%% file: main.tex
\documentclass[conference]{IEEEtran}

\IEEEoverridecommandlockouts

\usepackage{cite}
\usepackage{amsmath,amssymb,amsfonts}
\usepackage{algorithmic}
\usepackage{graphicx}
\usepackage{textcomp}
\usepackage{microtype}

\def\BibTeX{{\rm B\kern-.05em{\sc i\kern-.025em b}\kern-.08em
    T\kern-.1667em\lower.7ex\hbox{E}\kern-.125emX}}

\input{macros}

\begin{document}

\title{A Compiler Framework for Optimizing \\ Dynamic Parallelism on GPUs
}

\input{authors}

\maketitle

\input{sec/0-abstract}

\input{sec/1-introduction}

\input{sec/2-background}

\input{sec/3-thresholding}

\input{sec/4-coarsening}

\input{sec/5-aggregation}

\input{sec/6-framework}

\input{sec/6-methodology}

\input{sec/7-evaluation}

\input{sec/8-related}

\input{sec/9-conclusion}

\input{sec/appendix}

\balance
\bibliographystyle{IEEEtranS}
\bibliography{main}

\end{document}

%% file: macros.tex
\usepackage{balance}
\usepackage{multirow}
\usepackage{color}
\usepackage{stfloats}

%% file: authors.tex

\author{
    \IEEEauthorblockN{
        Mhd Ghaith Olabi\textsuperscript{1},
        Juan G\'{o}mez Luna\textsuperscript{2},
        Onur Mutlu\textsuperscript{2},
        Wen-mei Hwu\textsuperscript{3,4},
        Izzat El Hajj\textsuperscript{1}
    }
    \IEEEauthorblockA{
        \\\textit{
        \textsuperscript{1}American University of Beirut, Lebanon~~~~
        \textsuperscript{2}ETH Z\"{u}rich, Switzerland~~~~
        \textsuperscript{3}NVIDIA, USA
        }\\\textit{
        \textsuperscript{4}University of Illinois at Urbana-Champaign, USA
        }
    }
}

%% file: sec/0-abstract.tex
\begin{abstract}

Dynamic parallelism on GPUs allows GPU threads to dynamically launch other GPU threads.
It is useful in applications with nested parallelism, particularly where the amount of nested parallelism is irregular and cannot be predicted beforehand.
However, prior works have shown that dynamic parallelism may impose a high performance penalty when a large number of small grids are launched.
The large number of launches results in high launch latency due to congestion, and the small grid sizes result in hardware underutilization.

To address this issue, we propose a compiler framework for optimizing the use of dynamic parallelism in applications with nested parallelism.
The framework features three key optimizations: thresholding, coarsening, and aggregation.
Thresholding involves launching a grid dynamically only if the number of child threads exceeds some threshold, and serializing the child threads in the parent thread otherwise.
Coarsening involves executing the work of multiple thread blocks by a single coarsened block to amortize the common work across them.
Aggregation involves combining multiple child grids into a single aggregated grid.

Thresholding is sometimes applied manually by programmers in the context of dynamic parallelism.
We automate it in the compiler and discuss the challenges associated with doing so.
Coarsening is sometimes applied as an optimization in other contexts.
We propose to apply coarsening in the context of dynamic parallelism and automate it in the compiler as well.
Aggregation has been automated in the compiler by prior work.
We enhance aggregation by proposing a new aggregation technique that uses multi-block granularity.
We also integrate these three optimizations into an open-source compiler framework to simplify the process of optimizing dynamic parallelism code.

Our evaluation shows that our compiler framework improves the performance of applications with nested parallelism by a geometric mean of 43.0$\times$ over applications that use dynamic parallelism, 8.7$\times$ over applications that do not use dynamic parallelism, and 3.6$\times$ over applications that use dynamic parallelism with aggregation alone as proposed in prior work.

\end{abstract}

%% file: sec/1-introduction.tex
\section{Introduction}

Dynamic parallelism on GPUs allows threads running on the GPU to launch grids of threads to also run on the GPU.
It is useful for programming applications with nested parallelism, particularly where the amount of nested parallelism is irregular and cannot be predicted at the beginning of the computation.
An example of such an application is graph processing where a thread visiting a vertex may want to perform some work for each of its neighbors.
In this case, the \textit{parent} thread visiting the vertex may launch a grid with many \textit{child} threads, one for each neighbor, to work on the neighbors concurrently.

Prior work has shown that using dynamic parallelism in this way imposes a high performance penalty~\cite{wang2014char,wang2015dynamic,el2016klap}.
The key problem is that when a massive number of small grids are launched, the launch latency is high because of the congestion caused by the large number of launches, and the device is underutilized because the grids are small in size.
To address this issue, various hardware and software optimizations have been proposed to reduce the overhead of dynamic parallelism.

Hardware techniques that have been proposed for reducing the overhead of dynamic parallelism include adding thread blocks to existing grids rather than launching new grids~\cite{wang2015dynamic} or providing a hardware controller that advises programmers on whether a launch is profitable~\cite{tang2017controlled}.
Further hardware optimizations include locality-aware scheduling of parent and child grids~\cite{wang2016isca,tang2018quantifying}.
However, these techniques require hardware changes so they are not available on current GPUs~\cite{cuda_guide}, which motivates the need for software techniques.

One category of software techniques is to have threads in the parent grid perform the nested parallel work without performing a dynamic launch~\cite{yang2014cudanp,chen2015free}.
These techniques mitigate the overhead of dynamic parallelism by avoiding it entirely, but require parent threads to be on standby regardless of whether or not there is work available for them to do.
Another category of techniques is to consolidate or aggregate the child grids being launched by multiple parent threads into a single grid~\cite{li2015exploiting,li2015nested,wu2016compiler,el2016klap}.
These techniques mitigate the overhead of dynamic parallelism by reducing the number of grids launched, hence the congestion, and increasing the sizes of the grids to ensure better hardware utilization.

In this paper, we propose a compiler framework for optimizing the use of dynamic parallelism that features three key optimizations: thresholding, coarsening, and aggregation.
The first optimization, thresholding, involves launching a grid dynamically only if the number of child threads exceeds a certain threshold, and serializing the work in the parent thread otherwise.
This optimization reduces the number of grids launched, thereby reducing congestion, and ensures that only large grids are launched that properly utilize the hardware.
Some prior works assume that thresholding is applied manually by the programmer~\cite{li2015exploiting,li2015nested,wu2016compiler,tang2017controlled}.
However, manual application of thresholding complicates the launch code, requires code duplication, and hurts code readability.
We propose to automate the thresholding optimization in the compiler and discuss the challenges associated with doing so.

The second optimization, coarsening, involves combining multiple child thread blocks into a single one.
This optimization reduces the number of child thread blocks that need to be scheduled, and interacts with the aggregation optimization to amortize the overhead of aggregation across the work of multiple original child blocks.
While coarsening is a common optimization applied in various contexts~\cite{kirk2016programming,magni2014automatic,stawinoga2018predictable}, we propose to apply it in the context of dynamic parallelism and observe its benefit in combination with aggregation.

The third optimization, aggregation, is similar to prior works~\cite{li2015exploiting,li2015nested,wu2016compiler,el2016klap} that aggregate child grids launched by multiple parent threads into a single grid.
However, the granularity of aggregation in prior work has been limited to launches by parent threads in either the same warp, the same thread block, or the entire grid.
We further enhance aggregation by proposing a new aggregation technique that uses multi-block granularity.
In this technique, launches are aggregated across parent threads in a group of thread blocks as opposed to the two extremes used in prior work of a single thread block or the entire grid.

In summary, we make the following contributions:
\begin{itemize}
    \item We present a compiler transformation that automates thresholding for dynamic parallelism (Section~\ref{sec:compiler}).
    \item We propose to apply coarsening in the context of dynamic parallelism, and present a compiler transformation for doing so (Section~\ref{sec:coarsening}).
    \item We propose to apply aggregation at multi-block granularity and present a compiler transformation for doing so (Section~\ref{sec:aggregation}).
    \item We combine thresholding, coarsening, and aggregation in one open-source compiler framework (Section~\ref{sec:framework}).
\end{itemize}
Our evaluation (Section~\ref{sec:evaluation}) shows that our compiler framework improves the performance of applications with nested parallelism by a geometric mean of 43.0$\times$ over applications that use dynamic parallelism, 8.7$\times$ over applications that do not use dynamic parallelism, and 3.6$\times$ over applications that use dynamic parallelism with aggregation alone as proposed in prior work.

%% file: sec/2-background.tex
\section{Background}\label{sec:opt}

\subsection{Dynamic Parallelism}\label{sec:opt-background}

\begin{figure}[t]
    \centering
    \includegraphics[width=\columnwidth]{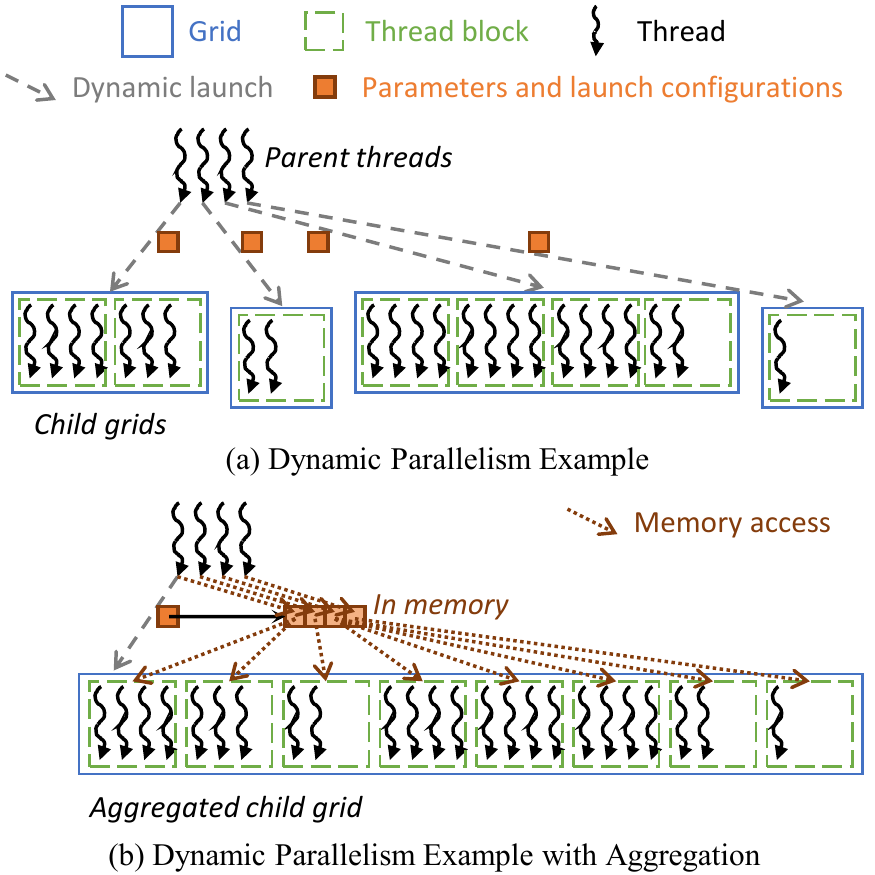}
    \caption{Background on Dynamic Parallelism}\label{fig:background}
    \vspace{-10pt}
\end{figure}

Fig.~\ref{fig:background}(a) shows an example of how dynamic parallelism can be used in practice.
In this example, parent threads executing on the GPU each discover some nested work that can be parallelized.
Each parent thread launches a child grid to perform the nested work in parallel, and each parent thread provides its child grid with a different set of launch configurations and parameters.
The amount of nested work may vary across threads.
Hence, the child grids have different sizes.
One source of inefficiency that may arise when using dynamic parallelism in this way is that a massive number of child grids may be launched, and many of them may be small in size~\cite{wang2014char}.
In this case, the large number of child grid launches causes congestion, and the small size of the child grids causes the device to be underutilized.

\subsection{Aggregation}\label{sec:opt-aggregation}

Aggregation is an optimization that consolidates or aggregates child grids that are launched by multiple parent threads into a single grid.
Various works~\cite{li2015exploiting,li2015nested,wu2016compiler,el2016klap} apply this optimization either manually or in the compiler. 
Fig.~\ref{fig:background}(b) shows how aggregation can be applied to the example in Fig.~\ref{fig:background}(a).
In this example, the parent threads coordinate to find the cumulative size of all their child grids in order to launch a single aggregated grid.
In prior work, the scope of coordination has been across parent threads in the same warp, block, or grid.
If the scope is across a warp or a block, one of the participating parent threads launches the aggregated grid on behalf of the others.
If the scope is across a grid, the aggregated grid is launched from the host.
The scope of coordination is referred to as the \emph{aggregation granularity}.

In the original code, each parent thread may provide different parameters and launch configurations to its child grid.
However, in the transformed code, only one set of parameters and launch configurations can be provided.
For this reason, before launching the aggregated grid, the parent threads each store their original parameters and launch configurations in memory, and a pointer to this memory is passed to the aggregated grid.
The child threads must then identify who their original parent thread is in order for them to load the right parameters and configurations from memory.
To do so, each child thread block executes a search operation.
The work done by the parent threads to identify the size of the aggregated grid and store their individual parameters and launch configurations in memory is referred to as the \emph{aggregation logic}. 
The work done by the child threads to identify their original parent thread and load their original parameters and configurations from memory is referred to as the \emph{disaggregation logic}.

The advantage of aggregation is that it reduces the number of child grids launched, thereby reducing congestion, and it increases the sizes of the child grids to ensure better hardware utilization.
The disadvantage of aggregation is that it incurs overhead due to the aggregation and disaggregation logic, and it delays the child grid launches until all parent threads are ready to launch.
The choice of aggregation granularity (warp, block, or grid) involves making a trade-off between these advantages and disadvantages.
Using a larger granularity reduces congestion and increases utilization, but incurs higher overhead from the aggregation and disaggregation logic and delays the child grids longer before launching them.

%% file: sec/3-thresholding.tex
\section{Thresholding}\label{sec:compiler}

\subsection{Optimization Overview}\label{sec:opt-thresholding}

\begin{figure}[h]
    \centering
    \includegraphics[width=\columnwidth]{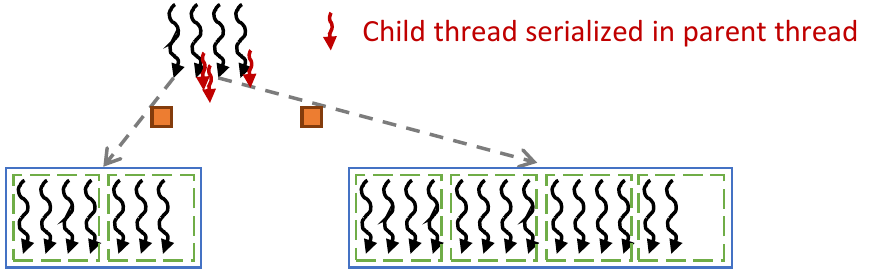}
    \caption{Dynamic Parallelism Example with Thresholding}\label{fig:thresholding}
\end{figure}

Thresholding is an optimization where a child grid is only launched dynamically if the number of child threads exceeds a certain \emph{threshold}.
Otherwise, the child threads are executed sequentially by the parent thread.
Fig.~\ref{fig:thresholding} illustrates how thresholding can be applied to the example in Fig.~\ref{fig:background}(a).
In this example, two of the parent threads have a small number of child threads.
The benefit gained from parallelizing these child threads is unlikely to be worth the launch overhead.
For this reason, the parent threads instead execute the work of the child threads sequentially.

Thresholding is commonly applied manually by programmers~\cite{li2015exploiting,li2015nested,wu2016compiler,tang2017controlled}.
However, manual application of thresholding complicates the launch code, requires code duplication, and hurts code readability.
For this reason, we propose to automate thresholding via a compiler transformation (Section~\ref{sec:thresholding-transform}) and discuss the challenges associated with doing so (Sections~\ref{sec:non-serializable} and~\ref{sec:identify-threads}).

\subsection{Code Transformation}\label{sec:thresholding-transform}

\begin{figure}
    \centering
    \includegraphics[width=\columnwidth]{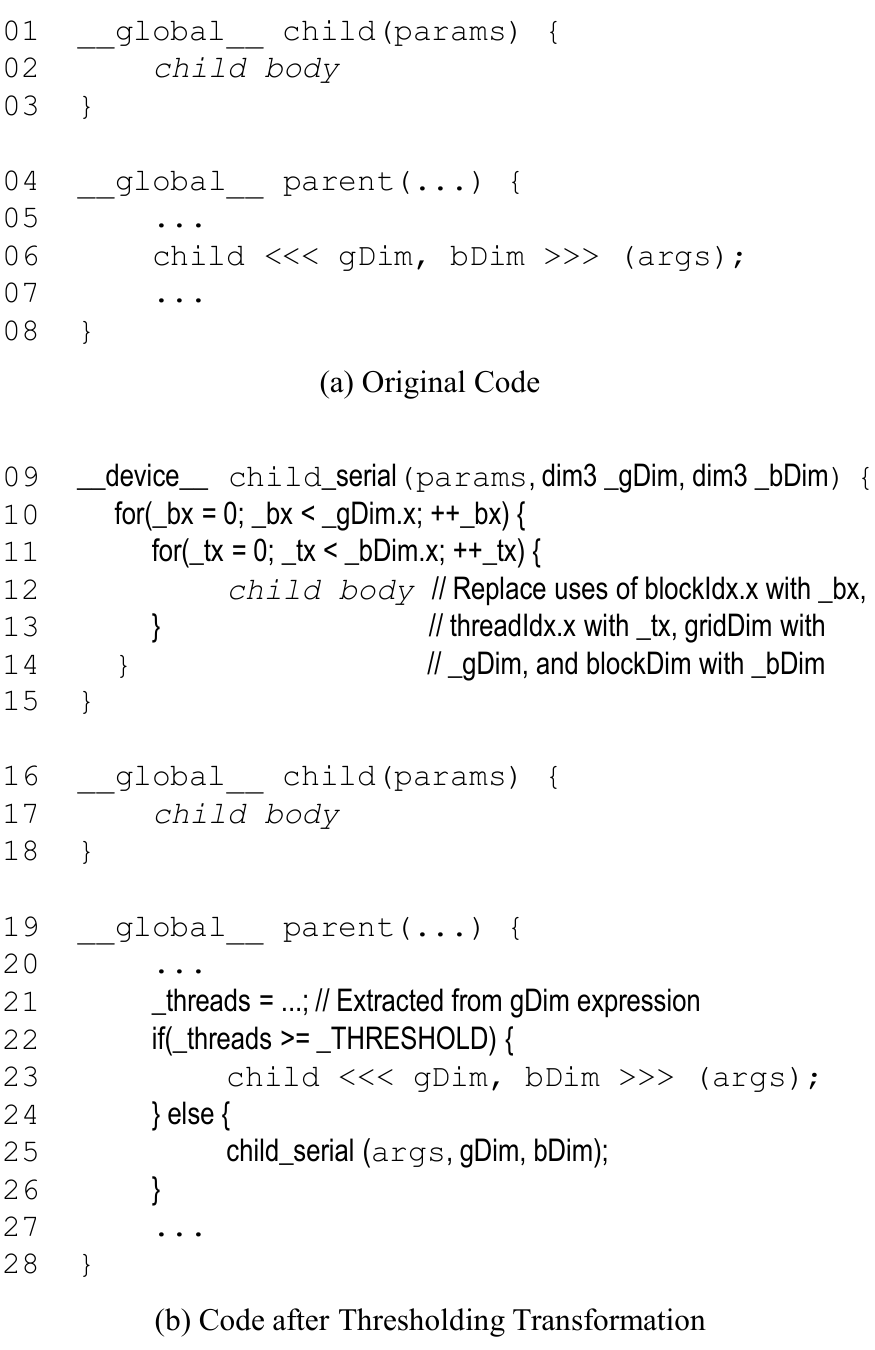}
    \caption{Thresholding Code Transformation Example}\label{fig:serialization-thresholding}
\end{figure}

Fig.~\ref{fig:serialization-thresholding} shows an example of how our compiler applies the thresholding transformation.
The original code in Fig.~\ref{fig:serialization-thresholding}(a) consists of a parent kernel (lines 04-08) and a child kernel (lines 01-03).
The parent kernel calls the child kernel using dynamic parallelism (line 06) and configures it with a grid dimension \texttt{gDim} and a block dimension \texttt{bDim}.

Fig.~\ref{fig:serialization-thresholding}(b) shows the code after the thresholding transformation is applied.
The transformation consists of two key parts: constructing a serial version of the child to be executed by the parent thread (lines 09-15) and applying a threshold to either perform the launch or call the serial version (lines 21-26).

To construct the serial version, the child kernel is replicated and its attribute is changed to \texttt{\_\_device\_\_} so that it becomes a device function (line 09).
Two parameters are appended to the parameter list: \texttt{\_gDim}, which represents the original grid dimension of the parallel version, and \texttt{\_bDim}, which represents the original block dimension.
Loops are inserted around the child body (lines 10-11) to serialize the child threads.
The first loop (line 10) iterates over the child thread blocks while the second loop (line 11) iterates over the child threads in each child block, using the bounds passed as parameters.
Finally, all uses of the reserved index and dimension variables are replaced with the corresponding loop indices and bounds.
The example shows a 1-dimensional child kernel for simplicity, however, if the child kernel is multi-dimensional, loops would be inserted for each dimension.

To apply the threshold, the number of child threads is first identified and stored in the \texttt{\_threads} variable (line 21).
Section~\ref{sec:identify-threads} discusses how the number of child threads is identified.
Next, an if-statement is inserted around the child kernel call (line 22) to ensure that the dynamic launch is performed only if the number of child threads is greater than or equal to \texttt{\_THRESHOLD}.
Here, \texttt{\_THRESHOLD} is a macro variable that can be overridden at compile time for tuning purposes.
If the threshold is not met, then the device function implementing the serial version is called instead (line 25), thereby serializing the child work in the parent thread.

\subsection{Non-Transformable Kernels}\label{sec:non-serializable}

Not all child kernels are amenable to the kind of transformation described in Section~\ref{sec:thresholding-transform}.
In particular, there are two kinds of child kernels that we do not transform: (1) child kernels that perform barrier synchronization across threads via \texttt{\_\_syncthreads()} or warp-level primitives, and (2) child kernels that use shared memory.

For child kernels that perform barrier synchronization, serializing GPU threads while supporting such synchronization has been done in the literature~\cite{stratton2008mcuda,kim2015locality,karrenberg2012improving,kim2014multi,el2014dynamic}.
The key strategy is to divide the code into regions separated by the barriers and insert loops around each region.
However, these techniques target serializing multiple GPU threads in one CPU thread.
Extending them to serialize multiple GPU threads in one GPU thread is not practical for two reasons.
The first reason is that these techniques perform scalar expansion of all local variables to preserve the state of all threads across barriers.
Such a scalar expansion on the GPU would convert all register accesses to memory accesses which would be prohibitively expensive.
The second reason for not serializing child threads that perform barrier synchronization is that code that includes barrier synchronizations often implements a parallel algorithm that is not efficient when serialized.
For example, a parallel reduction operation uses a reduction tree and leverages barriers to synchronize between levels of the tree.
However, reduction trees are not an efficient way to perform sequential reductions. 
It is more efficient to use a simple reduction loop.
In this case, it is better to let the programmer apply the thresholding optimization manually because the best sequential and parallel algorithms are different.

For child kernels that use shared memory, we do not construct a serial version of the kernel because every parent thread would require as much shared memory as an entire child block which would make the shared memory requirements of a parent block too high.
Besides, kernels that use shared memory most often use \texttt{\_\_syncthreads()} to coordinate access to shared memory across threads.
Hence, these kernels will most likely not be transformable anyway for the reasons related to barrier synchronization previously mentioned.

\subsection{Identifying the Number of Child Threads}\label{sec:identify-threads}

The transformation described in Section~\ref{sec:thresholding-transform} needs to identify the desired number of child threads in order to compare it with the threshold.
Identifying the desired number of child threads is challenging because this information is not what the programmer provides in the kernel call.
Instead, the programmer provides the grid dimension (number of blocks) and the block dimension (number of threads per block).
The programmer uses the desired number of child threads in calculating the grid dimension.

One way to identify the desired number of child threads is to multiply the grid dimension with the block dimension.
However, this approach gives the total number of threads in the child grid including threads that will not be used.
This value is not a representative value to compare with the threshold.
For example, consider a kernel with a nested kernel call where the child block dimension is configured to 1024 threads.
One of the parent threads identifies 2 units of nested parallel work to be processed by the child.
This parent thread will configure its child kernel with 1 block.
In this case, multiplying the grid dimension with the block dimension gives 1024 total threads which is much larger than the actual number of threads desired (2 threads).
Ideally, the value that should be compared to the threshold is 2, not 1024.
Hence, multiplying the grid dimension with the block dimension is not a good approach.

The approach we use to identify the desired number of child threads is based on the observation that programmers usually calculate the grid dimension as a ceiling-division of the desired number of threads by the block dimension.
There are arbitrary ways in which a ceiling-division can be expressed so it is not possible to have a static analysis that always determines the desired number of threads with certainty.
Instead, we identify the most common patterns used by programmers to calculate ceiling-division and employ a simple static analysis based on these patterns.

\begin{figure}[h]
    \centering
    \includegraphics[width=\columnwidth]{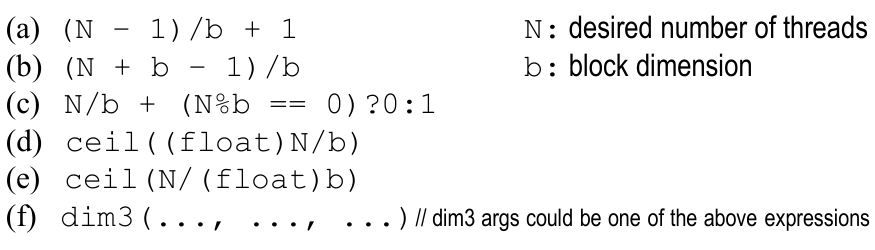}
    \caption{Common Expressions for Calculating the Grid Dimension}\label{fig:ceiling}
\end{figure}

Fig.~\ref{fig:ceiling} shows common expressions that programmers write for calculating the grid dimension using ceiling-division.
Options (a)-(c) use integer arithmetic, while options (d)-(e) convert to floating point and use the \texttt{ceil} function.
Option (f) is used in multi-dimensional blocks, where the operands to the \texttt{dim3} constructor could each be an expression that resembles options (a)-(e).
For all options, the expression may be expressed as a whole, or it may be expressed in parts where subexpressions are stored in intermediate variables.
Note that \texttt{N} and \texttt{b} can be arbitrary expressions.

We observe from the examples in Fig.~\ref{fig:ceiling} that \texttt{N} is usually in the subexpression on the left hand side of the division.
Moreover, the subexpression containing \texttt{N} may also contain constants such as 1 or \texttt{b} (which is usually a constant).
Based on this observation, our analysis pass looks for a division operation, takes the subexpression on the left hand side, and removes additions and subtractions of constants, considering the remaining subexpression as the desired number of threads.
This analysis is heuristic by nature and is not guaranteed to find the true desired number of threads.
However, using a heuristic is acceptable in this context because the result will only be used to choose whether to serialize or parallelize the work.
This choice does not impact program correctness in any way.

The subexpression that is found is assigned to \texttt{\_threads} in Fig.~\ref{fig:serialization-thresholding} on line 21.
The occurrence of the subexpression in \texttt{gDim} is then replaced with \texttt{\_threads} to ensure that the expression is not duplicated in the code just in case the expression has side effects.

%% file: sec/4-coarsening.tex
\section{Coarsening}\label{sec:coarsening}

\subsection{Optimization Overview}\label{sec:opt-coarsening}

Coarsening is an optimization where the work of multiple thread blocks in the original code is assigned to a single thread block.
The number of original thread blocks assigned to each coarsened thread block is referred to as the \textit{coarsening factor}.
When the GPU hardware is oversubscribed with more thread blocks than it can accommodate simultaneously, the hardware serializes these thread blocks, scheduling a new one whenever an old one has completed.
There are multiple advantages of serializing these thread blocks in the code rather than letting the hardware do it.
First, it reduces the number of thread blocks that need to be scheduled, and allows some warps in the coarsened block to proceed to executing the work of the next original block before other warps have completed their part in the previous original block.
Second, if there is common work across the original thread blocks, that work can be factored out and executed once by the coarsened thread block, allowing its cost to get amortized.
The disadvantage of coarsening is that it reduces parallelism, thereby underutilizing the device if the coarsening factor is too high.

Coarsening as an optimization is often applied by programmers in many different contexts~\cite{kirk2016programming}.
Prior works have also applied coarsening in the compiler~\cite{magni2014automatic,stawinoga2018predictable}, but not in the context of dynamic parallelism.
We propose to apply coarsening in the context of dynamic parallelism, and automate its application via a compiler transformation.

\begin{figure}[b]
    \centering
    \includegraphics[width=\columnwidth]{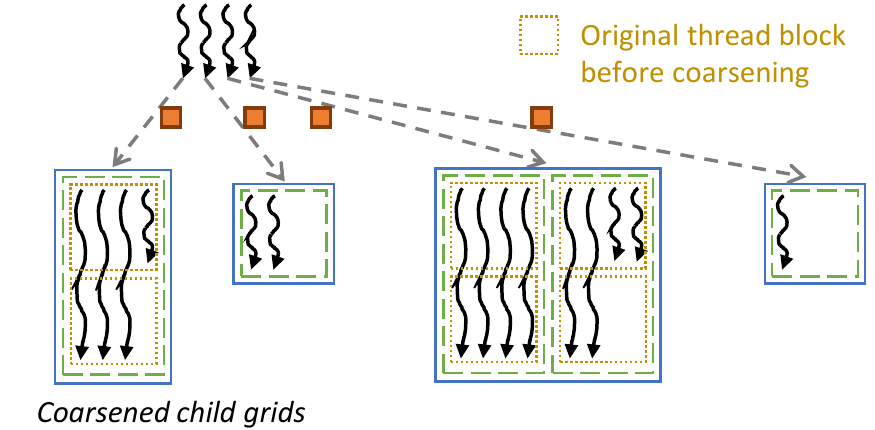}
    \caption{Dynamic Parallelism Example with Coarsening}\label{fig:coarsening}
\end{figure}

Fig.~\ref{fig:coarsening} illustrates how coarsening can be applied to the child thread blocks in the example in Fig.~\ref{fig:background}(a).
In this example, each coarsened child thread block in the transformed code executes the work of two child thread blocks in the original code.
The advantage of applying coarsening in the context of dynamic parallelism is that it reduces the number of child thread blocks that need to be scheduled.
More importantly, when applied before aggregation, coarsening also has the advantage of providing child thread blocks that do more work per block, which makes them more capable of amortizing the disaggregation logic overhead.

\subsection{Code Transformation}\label{sec:coarsening-transform}

\begin{figure}
    \centering
    \includegraphics[width=\columnwidth]{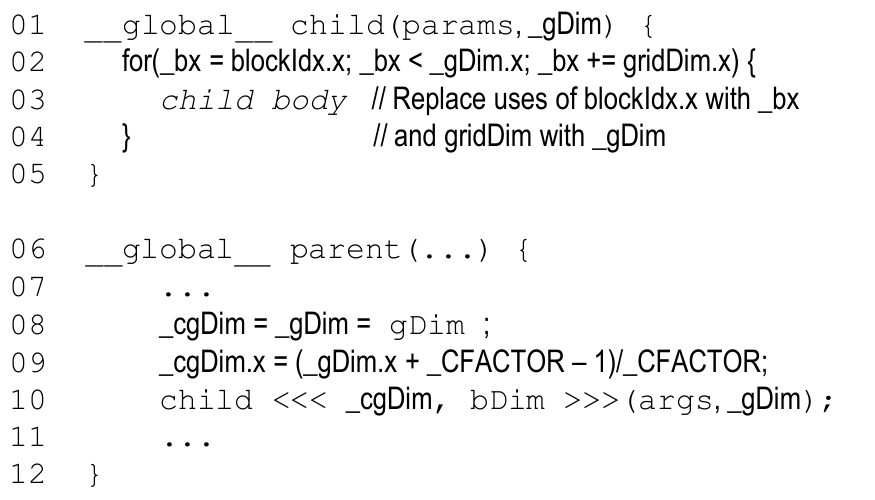}
    \caption{Coarsening Code Transformation Example}\label{fig:coarsening-code}
\end{figure}

Fig.~\ref{fig:coarsening-code} shows an example of how our compiler applies the coarsening transformation to the original code in Fig.~\ref{fig:serialization-thresholding}(a).
The transformation consists of two key parts: coarsening the child kernel (lines 01-05) and modifying the launch configurations to launch the coarsened child (lines 08-10).

To coarsen the child kernel, a parameter \texttt{\_gDim} is appended to the parameter list (line 01) which represents the original grid dimension without coarsening.
A \textit{coarsening loop} is inserted (line 02) that iterates over the work of the original child thread blocks assigned to the coarsened block.
Uses of the reserved index and dimension variables are replaced with the corresponding loop indices and bounds.
The example shows coarsening in one dimension only for simplicity, however, if the child grid is multi-dimensional, loops would be inserted for each dimension.

To modify the launch configurations to launch the coarsened child, the original grid dimension \texttt{gDim} is stored in a variable \texttt{\_gDim} (line 08).
The value is also copied to \texttt{\_cgDim}, which represents the coarsened grid dimension.
The x-dimension of the coarsened grid dimension \texttt{\_cgDim} is then ceiling-divided by the coarsening factor \texttt{\_CFACTOR} (line 09).
Here, \texttt{\_CFACTOR} is a macro variable that can be overridden at compile time for tuning purposes.
Again, the example shows coarsening in one dimension for simplicity.
Finally, the child kernel is configured with the coarsened grid dimension, and the original grid dimension is passed as a parameter (line 10).

%% file: sec/5-aggregation.tex
\section{Aggregation}\label{sec:aggregation}

Aggregation has been proposed by prior work~\cite{li2015exploiting,li2015nested,wu2016compiler,el2016klap} and has been described in Section~\ref{sec:opt-aggregation}.
In this paper, we propose a new aggregation granularity, namely, multi-block granularity (Section~\ref{sec:multi-block-agg}).
We also propose to apply an aggregation threshold to optimize aggregation at warp and block granularity (Section~\ref{sec:agg-threshold}).

\subsection{Multi-block Granularity Aggregation}\label{sec:multi-block-agg}

Recall from Section~\ref{sec:opt-aggregation} that prior work has performed aggregation at warp, block, and grid granularity.
Using a larger granularity reduces congestion and increases hardware utilization, but incurs higher overhead from the aggregation and disaggregation logic and delays the child grids longer before launching them.
The choice of aggregation granularity presents a trade-off between these performance factors.
However, there is a wide gap between the block and grid granularity aggregation schemes that leaves a large part of the trade-off space unexplored. 
A parent thread block typically has hundreds to up to 1,024 parent threads, whereas a parent grid can have many thousands to millions of parent threads.
To better explore the trade-off space, we propose an intermediate granularity between block and grid granularity, which is multi-block granularity aggregation.

In multi-block granularity aggregation, we divide a parent grid into groups of blocks, each group having a fixed number of blocks.
The threads within the same group of blocks collaborate to aggregate their child grids into a single aggregated grid.
In prior work~\cite{el2016klap}, the aggregation logic involves a scan operation on the original grid dimensions and a max operation on the block dimension to identify the aggregated grid configuration.
It also involves a barrier synchronization to wait for all participating threads to store their individual configurations and arguments to memory before the aggregated launch is performed.
Since we cannot synchronize across multiple thread blocks, we perform the scan and max operations for multi-block granularity using atomic operations similar to what is done at grid granularity in prior work.
As for the barrier synchronization, we replace it with a group-wide counter that is atomically incremented by each thread block when it finishes.
The last thread block in the group to increment the counter performs the aggregated launch.

\begin{figure}[t]
    \centering
    \includegraphics[width=\columnwidth]{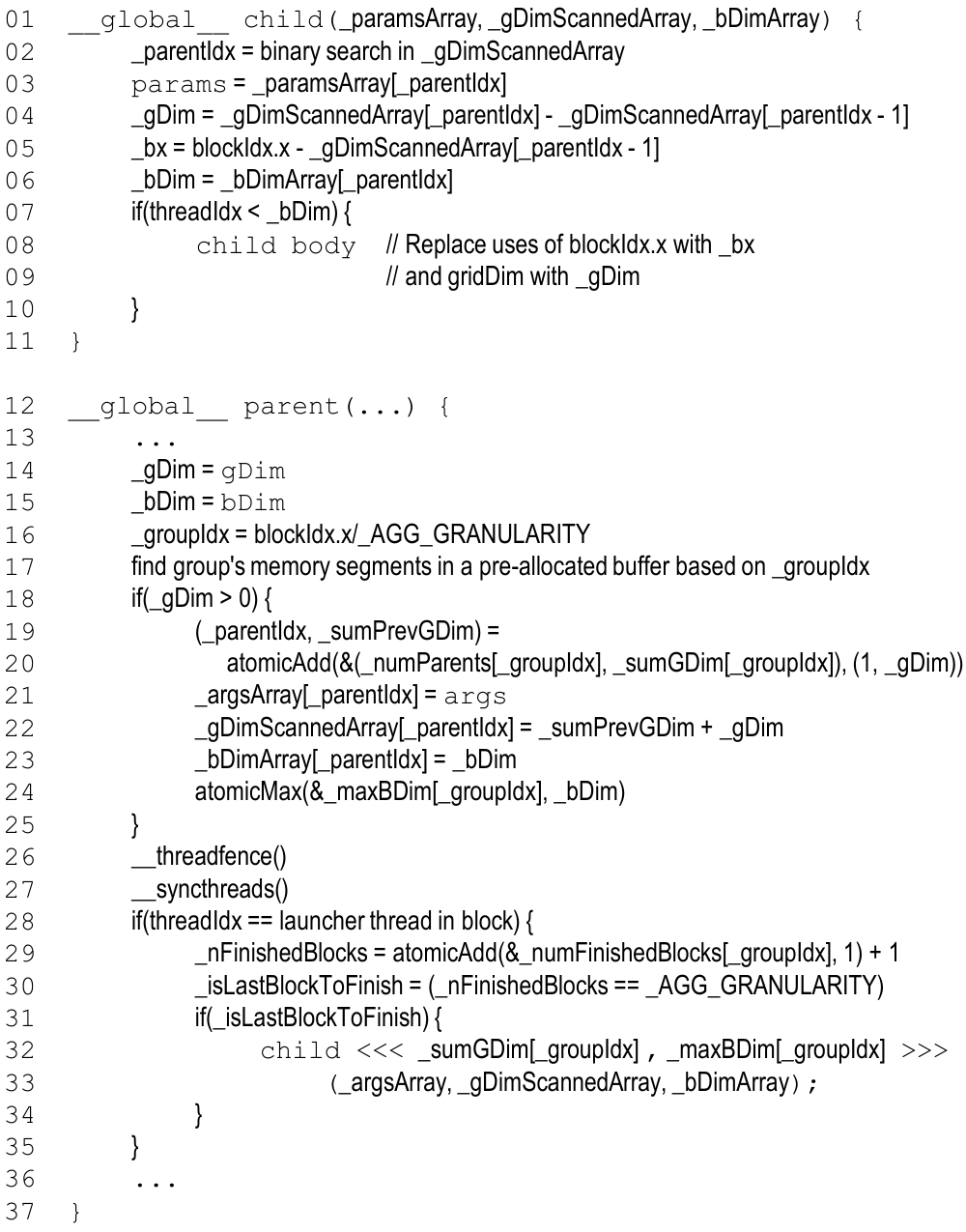}
    \caption{Multi-block Granularity Aggregation Code Transformation Example}\label{fig:aggregation-code}
\end{figure}

Fig.~\ref{fig:aggregation-code} shows an example of how our compiler applies the multi-block granularity aggregation transformation to the original code in Fig.~\ref{fig:serialization-thresholding}(a).
The transformation consists of two parts: the aggregation logic in the parent kernel (lines 14-35) and the disaggregation logic in the child kernel (lines 01-11).

For the aggregation logic in the parent kernel, we save the \texttt{gDim} and \texttt{bDim} expressions in temporary variables to avoid recomputing them every time we use them in case they have side effects (lines 14-15).
We then identify the group that the thread block belongs to (line 16) and find the group's memory segment in a pre-allocated memory buffer (line 17).
This memory segment is used to store the configurations and arguments that threads in the group pass to their children.
Next, each thread that launches a child grid (has a non-zero grid dimension) atomically increments two global counters simultaneously (lines 19-20): (1) \texttt{\_numParents} to assign an index to the parent thread so that the thread knows where to
store its arguments and configuration, and (2) \texttt{\_sumGDim} to find the total number of child blocks of prior parent threads which we use to initialize the scanned array of grid dimensions.
The two global counters are incremented simultaneously by treating them as a single 64-bit integer.
Each thread then stores its arguments, scanned grid dimension, and block dimension to memory so they can be passed to the child grid (lines 21-23).
Each thread also performs an atomic operation to find the maximum block dimension (line 24).

After each thread writes its configuration and arguments to global memory, it performs a fence operation to ensure that the configuration and arguments are visible to its child blocks (line 26).
This fence is not necessary in prior work because the visibility of the data is ensured either by the semantics of the dynamic kernel call (for warp and block granularity) or by grid termination (for grid granularity).
However, it is necessary here because the launch may be performed by a different thread block than the one that writes the data. Hence, the writing thread block must ensure that its data is visible in global memory before notifying the other blocks that it is ready.
A local barrier (line 27) is also needed to ensure that all threads in the block finish storing their data before notifying.

Finally, it is time to perform the aggregation launch.
One thread in the block (line 28) atomically increments the group-wide counter (line 29) and checks if its block is the last block in the group to finish (line 30-31).
If so, the thread launches the aggregated grid, configuring it with the sum of all grid dimensions and the maximum of all block dimensions, and passing pointers to the memory arrays containing the original arguments and configurations (lines 32-33).

For the disaggregation logic in the child kernel (lines 01-11), the code remains largely the same as prior work~\cite{el2016klap}.
Each child block performs a binary search through the scanned grid dimension array to identify its original parent thread (line 02).
It then loads its parameters and configuration (lines 04-06) and performs the work of the child kernel based on them (lines 07-10).

Note that besides filling the gap in the trade-off space between block and grid granularity aggregation, multi-block granularity aggregation has another advantage over grid granularity aggregation in particular.
Grid granularity aggregation requires the CPU to be involved in the aggregated launch whereas the multi-block granularity aggregation runs entirely on the GPU, which frees the CPU to perform other tasks.
Hence, multi-block granularity aggregation is more compatible with the asynchronous semantics of kernel calls.
If the CPU is needed for performing other tasks, multi-block granularity aggregation may be better for overall execution time even if grid granularity aggregation has better kernel time.

\subsection{Aggregation Threshold}\label{sec:agg-threshold}

When thresholding is applied before aggregation, the number of original child grids that participate in the aggregated grid may be substantially reduced.
If there is an insufficient number of original child grids participating in the aggregated grid, the benefit of aggregation may not be worth its overhead.
To address this issue, we enhance aggregation with another optimization that applies an \textit{aggregation threshold}.
The aggregation logic is preceded by an operation to count the number of participating parent threads.
If the number of participating parent threads does not meet a certain threshold, the child grids are launched normally by their parent threads instead of being aggregated.
Therefore, a child grid may be executed in one of three ways: it may be serialized within its parent, launched directly by its parent, or launched as part of an aggregated grid.
Since applying an aggregation threshold requires parent threads to synchronize to count the number of participating threads, it can only be applied at warp and block granularity where barrier synchronization across threads is possible.

%% file: sec/6-framework.tex
\section{Compiler Framework}\label{sec:framework}

We integrate our three optimizations -- thresholding, coarsening, and aggregation -- into a single compiler framework.
For separation of concerns, each optimization is implemented as a separate source-to-source transformation pass that takes a CUDA \texttt{.cu} file and generates a \texttt{.cu} file.
The transformations are independent, meaning that any combination of them could be applied in any order while generating correct code.

\begin{figure}
    \centering
    \includegraphics[width=\columnwidth]{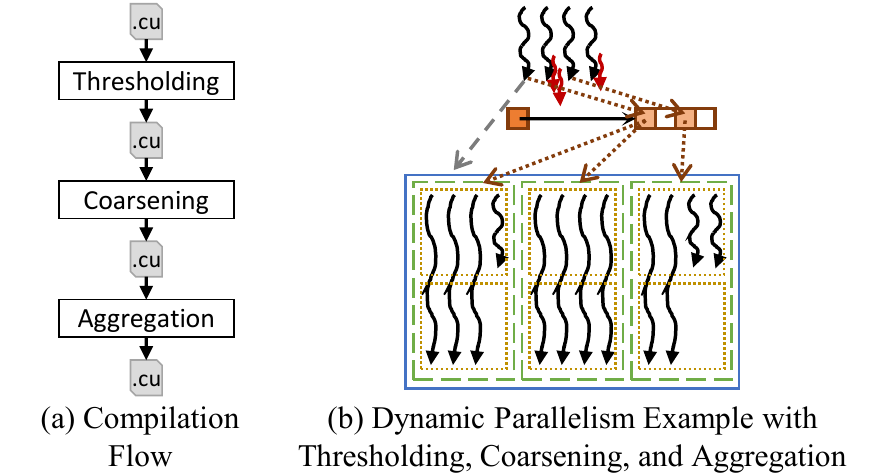}
    \caption{Combining the three optimizations}\label{fig:framework}
\end{figure}

Although the optimizations can be applied in any order, we apply them in the following order: thresholding, coarsening, then aggregation, as shown in Fig.~\ref{fig:framework}(a).
Thresholding is applied before coarsening because coarsening manipulates the grid dimension which makes it harder to extract the number of threads that needs to be compared to the threshold.
Thresholding is applied before aggregation because aggregation combines small grids with large grids into a single aggregated grid.
It is more difficult to isolate small grids and serialize them after they have been aggregated into larger ones.
Coarsening is applied before aggregation because the disaggregation logic should be outside the coarsening loop so that it can be amortized across multiple original child blocks.

Figure~\ref{fig:framework}(b) illustrates the impact of combining all three optimizations on the example in Fig.~\ref{fig:background}(a).
In this example, two of the parent threads have small child grids so the work of the child grids is serialized in the parent threads by the thresholding optimization.
The remaining two parent threads have large child grids so they collaborate to aggregate their grids and perform a single launch.
Each child block in the aggregated grid searches for its parent thread and obtains the corresponding parameters and configuration from memory.
The child block then executes the work of multiple original child blocks because it was coarsened prior to aggregation.

The compiler transformations are implemented as source-to-source transformation passes in Clang~\cite{lattner2004llvm}.
The thresholding and coarsening transformations and their supporting analyses are implemented from scratch.
The aggregation transformations are implemented by modifying the implementations provided by one of the prior works~\cite{el2016klap}.
We leverage this work because it is open source, but our techniques can also be applied to any of the prior works that perform aggregation~\cite{el2016klap,wu2016compiler}.

%% file: sec/6-methodology.tex
\section{Methodology}\label{sec:methodology}

We evaluate our compiler framework on a system with a Volta V100 GPU~\cite{nvidia2017v100} with 32GB of device memory and an AMD EPYC 7551P CPU~\cite{amd} with 64GB of main memory.
Table~\ref{tab:benchmarks-and-datasets} shows the benchmarks and datasets used in our evaluation.
We set the pending launch count appropriately to avoid overflowing the launch buffer pool~\cite{cuda_guide}.
We compile the benchmarks with per-thread default streams enabled to ensure that launches from the same block are not bottlenecked on the same default stream.
We use larger datasets than prior work~\cite{el2016klap} does because we evaluate on a larger GPU.
However, our evaluation on smaller datasets shows similar trends.
Note that for TC, we use parts of the graphs in Table~\ref{tab:benchmarks-and-datasets} due to memory constraints.

\begin{table}[t]
    \centering
    \caption{Benchmarks and Datasets}\label{tab:benchmarks-and-datasets}
    \resizebox{\columnwidth}{!}{
        \input{fig/6-methodology/benchmarks-and-datasets}

    }
\end{table}

\begin{figure*}[b]
    \centering
    \includegraphics[width=\textwidth]{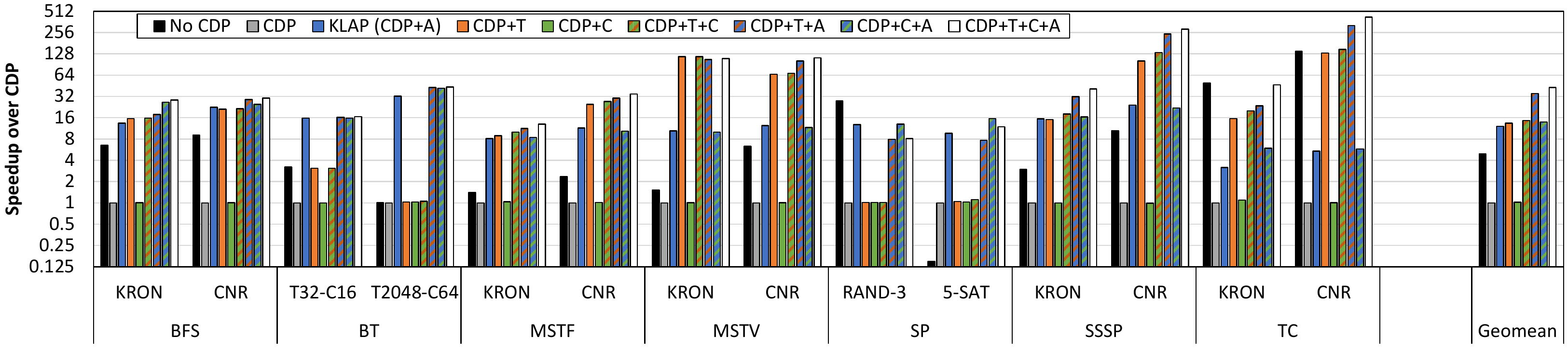}
    \caption{Performance (higher is better)
    }\label{fig:performance}
\end{figure*}

We compare our results to three different baselines.
The \emph{No CDP} versions of benchmarks are the original versions cited in Table~\ref{tab:benchmarks-and-datasets} that do not use CUDA Dynamic Parallelism.
The \emph{CDP} versions use CUDA Dynamic Parallelism and are obtained from prior work~\cite{el2016klap}.
The \emph{KLAP (CDP+A)} versions perform aggregation only and are also obtained from prior work~\cite{el2016klap}.
We use KLAP as a baseline from among prior works because it is open source and because we build on it in our compiler framework.

We report the performance of code versions generated by our compiler from the CDP version for multiple combinations of optimizations.
We indicate the combinations of optimizations applied as follows: \emph{T} for thresholding, \emph{C} for coarsening, and \emph{A} for aggregation.
The only exception is the TC benchmark with CDP+T because the original benchmark already applies dynamic parallelism with thresholding.

For each combination of optimizations, we tune the relevant parameters and report results for the best configuration.
The tuned parameters are the launch threshold, coarsening factor, and aggregation granularity.
The threshold is not tuned beyond the largest dynamic launch size to ensure that at least one dynamic launch is performed.
We use an exhaustive search to perform tuning
to show the maximum potential of the optimizations and to present a complete view of the design space.
However, such an exhaustive search is unnecessary in practice as we discuss in Section~\ref{sec:results-TA}.

To extract the breakdown of execution time in Section~\ref{sec:breakdown}, we incrementally deactivate portions of the code and calculate the time difference.
For benchmarks with multiple iterations, we report the time for the longest running iteration.
For BT, the aggregated \texttt{cudaMalloc} in the parent is considered part of the parent work because it is not affected by thresholding and coarsening.

%% file: fig/6-methodology/benchmarks-and-datasets.tex
\begin{tabular}{|l|l|l|}
    \hline
    \textbf{Benchmark} & \textbf{Description} & \textbf{Dataset Used} \\
    \hline
    BFS     & Breadth First Search \cite{Danalis2010Scalable}                  & KRON, CNR \\
    BT      & Bezier Tessellation \cite{nvidia2015nvidia}                   & T0032-C16, T2048-C64 \\
    MSTF    & Minimum Spanning Tree (find kernel) \cite{Burtscher2020Quantity}   & KRON, CNR \\
    MSTV    & Minimum Spanning Tree (verify kernel) \cite{Burtscher2020Quantity} & KRON, CNR \\
    SP      & Survey Propagation \cite{Burtscher2020Quantity}                   & RAND-3, 5-SAT \\
    SSSP    & Single Source Shortest Path \cite{Burtscher2020Quantity}          & KRON, CNR \\
    TC      & Triangle Counting \cite{2018_collaborative_cpu_gpu_tc}            & KRON, CNR \\
    \hline
    \multicolumn{3}{c}{~}\\
    \hline
    \textbf{Dataset} & \multicolumn{2}{|l|}{\textbf{Description}} \\
    \hline
    KRON        & \multicolumn{2}{|l|}{kron\_g500-simple-logn16, 65,536 vertices, 2,456,071 edges \cite{nr}} \\
    CNR         & \multicolumn{2}{|l|}{cnr-2000, 325,557 vertices, 2,738,969 edges \cite{BoVWFI}}  \\
    T0032-C16   & \multicolumn{2}{|l|}{Max Tessellation 32, Curvature: 16, Lines: 20,000 \cite{nvidia2015nvidia}} \\
    T2048-C64   & \multicolumn{2}{|l|}{Max Tessellation: 2048, Curvature: 64, Lines: 20,000 \cite{nvidia2015nvidia}} \\
    RAND-3      & \multicolumn{2}{|l|}{random-42000-10000-3, 10,000 literals \cite{Burtscher2020Quantity}} \\
    5-SAT       & \multicolumn{2}{|l|}{5-SATISFIABLE, 117,296 literals \cite{belov2014proceedings}} \\
    \hline
\end{tabular}

%% file: sec/7-evaluation.tex
\section{Evaluation}\label{sec:evaluation}

\subsection{Performance}

Fig.~\ref{fig:performance} shows the performance results for each benchmark and dataset with all combinations of optimizations applied.
Performance is reported as speedup over the CDP version.

Applying CDP alone leads to a performance degradation in almost all cases compared to not applying CDP.
Aggregation alone recovers from this degradation and substantially improves performance, where CDP+A is 12.1$\times$ faster than CDP and 2.4$\times$ faster than No CDP.
These observations are consistent with those made in prior works~\cite{li2015exploiting,li2015nested,el2016klap,wu2016compiler}, which note that the large number of launches causes congestion and aggregating the grids mitigates this overhead.

Thresholding alone gives substantial speedup, where CDP+T is 13.4$\times$ (geomean) faster than CDP.
Thresholding also gives speedup in the presence of aggregation, where CDP+T+A is 2.9$\times$ (geomean) faster than CDP+A, and CDP+T+C+A is also 3.1$\times$ (geomean) faster than CDP+C+A.
The incremental benefit of thresholding with aggregation is not as pronounced as without aggregation because the benefit of thresholding is reducing the number of launches which aggregation also does.
Nevertheless, the speedup is still significant.

Coarsening without aggregation gives modest speedup, where CDP+C is 1.01$\times$ (geomean) faster than CDP and CDP+T+C is 1.09$\times$ (geomean) faster than CDP+T.
On the other hand, coarsening with aggregation gives more speedup, where CDP+C+A is 1.16$\times$ (geomean) faster than CDP+A, and CDP+T+C+A is 1.22$\times$ (geomean) faster than CDP+T+A.
Notice how coarsening is synergistic with aggregation (its speedup in the presence of aggregation is greater than its speedup in the absence of aggregation).
The reason is that in the presence of aggregation, coarsening helps amortize the disaggregation logic across more work in the coarsened child block in contrast with much less work in each original child block.
Although the benefit of coarsening is much smaller than that of thresholding and aggregation, it is still substantial.
A speedup of 1.22$\times$ is significant considering that it comes from a compiler optimization that requires no programmer intervention or additional architecture support.

Ultimately, our compiler framework as a whole substantially improves performance with the three optimizations combined, where CDP+T+C+A is 43.0$\times$ (geomean) faster than CDP.
CDP+T+C+A is also 8.7$\times$ (geomean) faster than No CDP, showing that dynamic parallelism is a powerful programming feature when combined with the right optimizations.
Moreover, CDP+T+C+A is 3.6$\times$ (geomean) faster than KLAP (CDP+A), showing that our framework provides substantial speedup over prior works that perform aggregation alone~\cite{el2016klap}.

\subsection{Breakdown of Execution Time}\label{sec:breakdown}

\begin{figure*}[b]
    \centering
    \includegraphics[width=\textwidth]{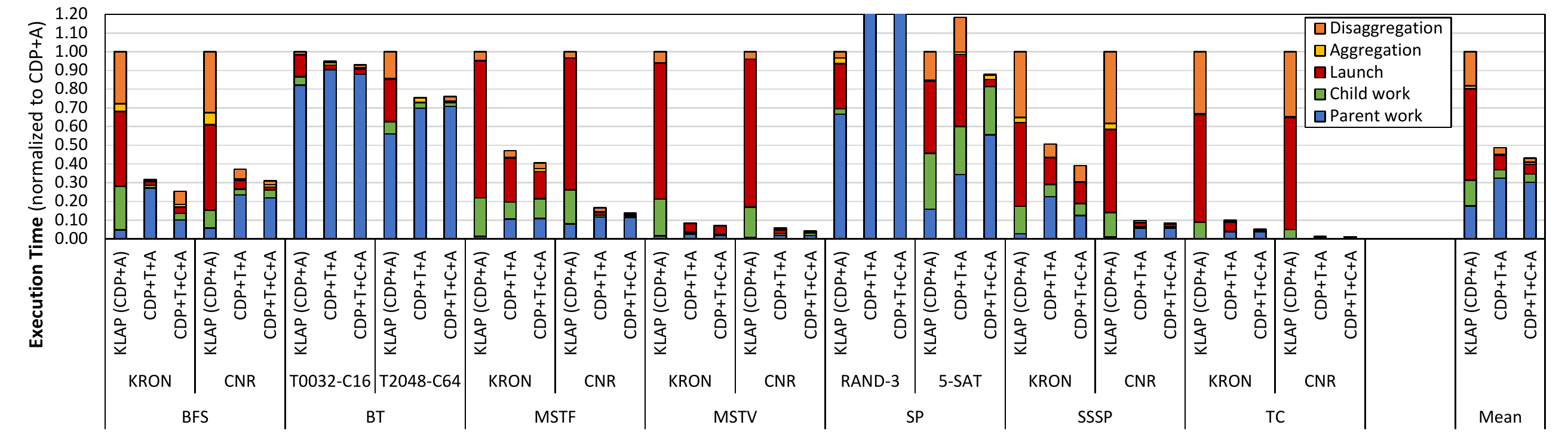}
    \caption{Breakdown of Execution Time (lower is better)}\label{fig:breakdown}
\end{figure*}

Fig.~\ref{fig:breakdown} shows how the execution time is spent for each benchmark and dataset.
We use KLAP (CDP+A) as baseline instead of CDP because prior work~\cite{el2016klap} has already shown how aggregation affects the execution time breakdown, and including CDP makes the figure illegible.
We compare with CDP+T+A and CDP+T+C+A to show the incremental impact of thresholding and coarsening, respectively.

The first observation is that thresholding increases parent work and decreases child work.
Thresholding serializes child work in parent threads so it is natural for parent threads to have more work to do and child threads to have less.

The second observation is that thresholding decreases the overhead from aggregation, launching, and disaggregation.
Thresholding results in fewer parent threads launching and fewer child threads being launched.
The aggregation overhead decreases because fewer parent threads participate, the launch overhead decreases because there are fewer launches, and the disaggregation overhead decreases because there are fewer child threads searching for their parents.

The third observation is that coarsening decreases the launch overhead.
Since coarsening reduces the number of child blocks that need to be scheduled, the launch overhead decreases.

The fourth observation is that coarsening decreases the disaggregation overhead.
Coarsening amortizes the disaggregation logic across multiple child thread blocks rather than having each child thread block perform its own disaggregation.

The final observation is that coarsening decreases parent work and increases child work for some benchmarks, namely BFS and SSSP.
This result is unintuitive because coarsening does not change the work that the parent does.
The reason is that for these two benchmarks, after coarsening is applied, the reduction in launch and disaggregation overhead results in a lower optimal threshold to be found.
The lower threshold results in more work being offloaded from parent to child.
This observation demonstrates an important interaction between thresholding and coarsening.

\subsection{Impact of Threshold and Aggregation Granularity}\label{sec:results-TA}

\begin{figure*}
    \centering
    \includegraphics[width=\textwidth]{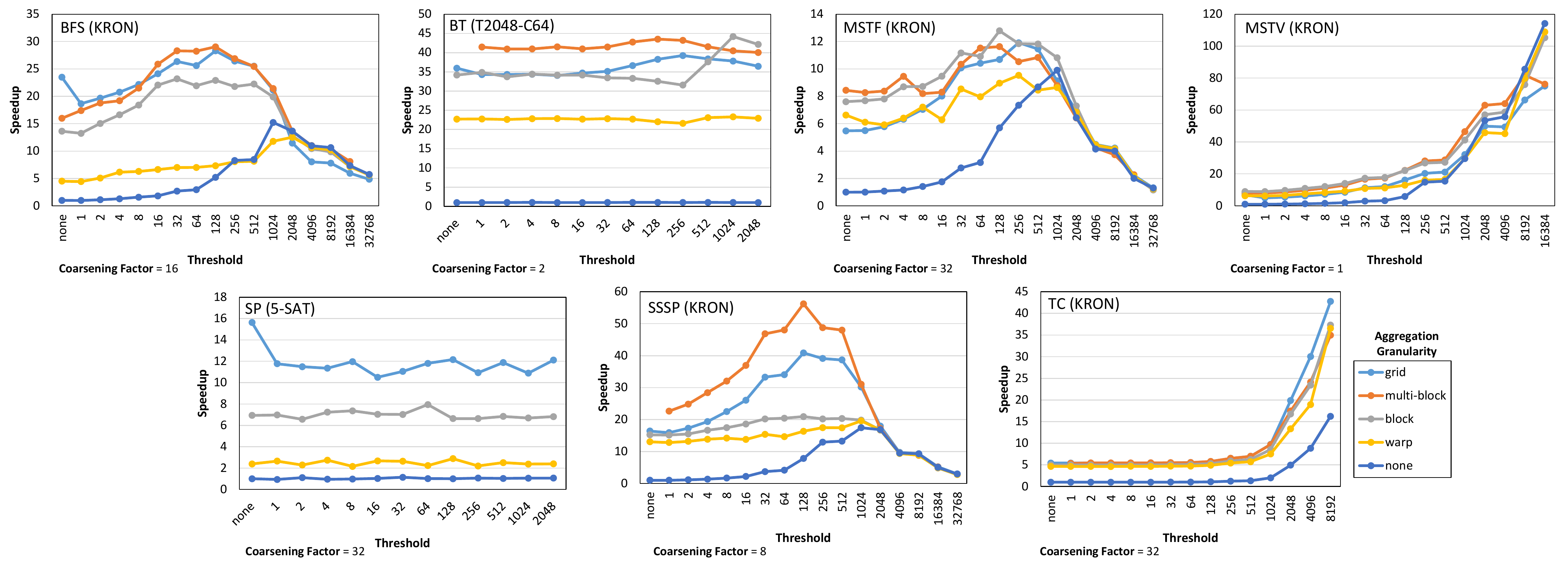}
    \caption{Impact of Threshold and Aggregation Granularity}\label{fig:sweep-ta-multiblock}
\end{figure*}

Fig.~\ref{fig:sweep-ta-multiblock} shows how performance varies for different threshold values and aggregation granularity while keeping the coarsening factor constant at the best coarsening factor found.
For space constraints, we only show the results for one dataset.

The first observation is that for most benchmarks (all except SP), as the threshold increases initially, performance also improves.
Increasing the threshold initially results in more of the small child grids getting serialized in their parent threads, which reduces the number of launches, hence the congestion, and avoids launching small grids that underutilize the device.

The second observation is that for some benchmarks (e.g., BFS, BT, MSTF, SSSP), increasing the threshold too much causes performance to degrade again.
Increasing the threshold too much results in large child grids getting serialized within their parent threads, which reduces parallelism and causes high control divergence.

The third observation is that different benchmarks perform best with different levels of aggregation granularity.
For example, SP and TC perform best with grid granularity, BFS and SSSP perform best with multi-block granularity, BT and MSTF perform best with block granularity, and MSTV performs best without aggregation.
The fact that multi-block granularity aggregation performs best in some of the cases reflects the importance of providing an intermediate granularity between grid and block granularity.

As mentioned in Section~\ref{sec:methodology}, we use an exhaustive search to perform tuning
to show the maximum potential of the optimizations and to present a complete view of the design space.
However, from our experience, such a broad search is unnecessary.
First, the best threshold is typically the one that allows approximately 6,000-8,000 child grid launches.
Second, performance is not very sensitive to the coarsening factor provided that it is sufficiently large ($>$8) so the coarsening factor does not need to be searched with a broad range or with high resolution.
Third, aggregation at warp granularity is never favorable.
With these observations in mind, users can typically find a combination of parameters that is very close to the best with less than ten runs.
Moreover, the compiler framework exposes these parameters in a configurable manner to make it easy for users to leverage off-the-shelf autotuners~\cite{ansel2014opentuner}.
Such tuning is worthwhile for kernels that run repeatedly on similar datasets.
On the other hand, if a user cannot afford to tune the kernel, it is not necessary for the user to find the best parameters to benefit from the optimizations.
Selecting a sub-optimal set of parameters would still yield a speedup, just not the maximum possible speedup.
For example, if we fix the threshold to 128 for all benchmarks and datasets, then CDP+T+C+A will have a geomean speedup of 1.9$\times$ over CDP+C+A, as opposed to 3.1$\times$ when the best threshold is used.
Hence, our optimizations are still useful even if the best parameters are not searched for or found.

\subsection{Workloads with Low Nested Parallelism}\label{sec:results-bad-datasets}

Dynamic parallelism is useful when the amount of nested parallelism is high such that the launch overhead is worth the parallel work extracted.
If the amount of nested parallelism is low for all parent threads, the benefit of dynamic parallelism is limited.
To demonstrate the impact of dynamic parallelism on applications with a small amount of nested parallelism, we evaluate the graph benchmarks on a road graph (USA-road-d.NY \cite{dimacs}).
The graph has 264,346 vertices, 730,100 edges, an average degree of 3, and a maximum degree of 8.
Hence, each vertex has a small number of outgoing edges so processing the graph has a small amount of nested parallelism.

\begin{figure}
    \centering
    \includegraphics[width=\columnwidth]{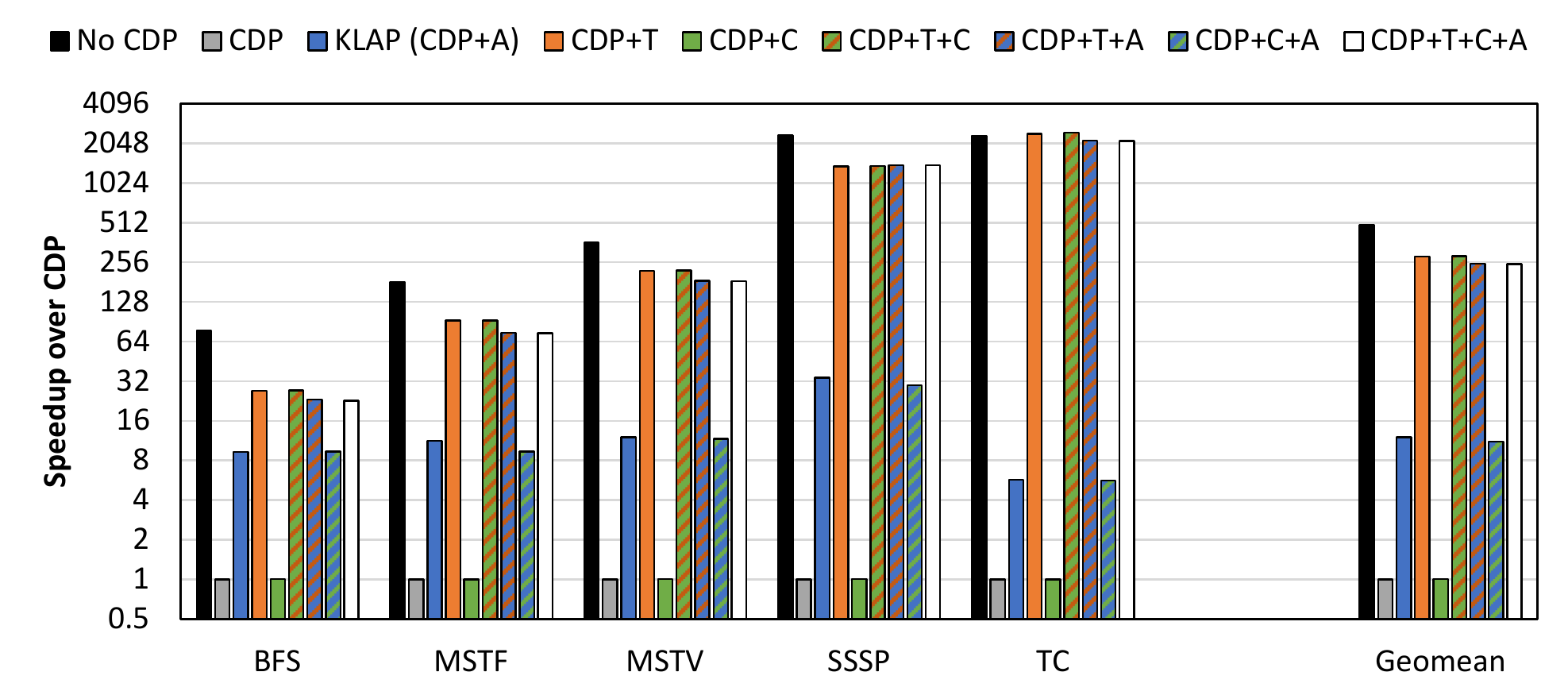}
    \caption{Performance of Graph Benchmarks on Road Graphs (higher is better)}\label{fig:speedup-road-graphs}
    \vspace{-10pt}
\end{figure}

Fig.~\ref{fig:speedup-road-graphs} shows the performance results for each graph benchmark on the road graph with all combinations of optimizations applied.
It is clear that CDP versions perform substantially more poorly on this graph relative to the No CDP versions.
Our proposed optimizations are able to recover much of the performance degradation, but not all of it.
In fact, in this experiment, we tune the threshold beyond the largest launch size such that CDP+T degenerates to serializing all child threads like No CDP.
However, CDP+T still cannot recover fully.
The reason is that the mere existence of a dynamic launch in the code, even if it is never executed, results in a performance degradation.
To verify this observation, we compare two kernels where the only difference is a dynamic launch that is guarded by a condition that is always false such that the launch is never performed.
Upon inspecting the assembly code of the two kernels, we observe that a large number of additional instructions are generated besides the instructions for performing the launch.
Upon profiling the execution of the two kernels, we observe that a large number of additional instructions are executed even though the launch is never performed.

Note that the SP benchmark on the RAND-3 dataset performs poorly in Fig.~\ref{fig:performance} also because the amount of nested parallelism is low (all child grids have fewer than 32 threads).
These results show the importance of being aware of the application and dataset when choosing whether or not to apply dynamic parallelism and its associated optimizations.

%% file: sec/8-related.tex
\section{Related Work}\label{sec:related}

To our knowledge, our work is the first to provide a compiler framework for optimizing dynamic parallelism code that combines the thresholding, coarsening, and aggregation optimizations together.

Several benchmarking efforts~\cite{wang2014char,ukidave2015nupar,harb2016characterizing} observe the inefficiencies of dynamic parallelism caused by high launch overhead and hardware underutilization, which has motivated various hardware and software optimizations.

Many hardware optimizations have been proposed for mitigating the overhead of dynamic parallelism.
Dynamic Thread Block Launch (DTBL)~\cite{wang2015dynamic,wang2016acceleration} proposes hardware support for lightweight dynamic launching of thread blocks rather than heavyweight dynamic launching of entire grids.
The dynamically launched thread blocks are essentially added to existing grids on the fly by the hardware.
LaPerm~\cite{wang2016isca} extends DTBL with a locality-aware scheduler.
SPAWN~\cite{tang2017controlled} is a hardware controller that advises programmers whether or not a dynamic launch is profitable.
LASER~\cite{tang2018quantifying} enhances dynamic parallelism with locality-aware scheduling.
Hardware optimizations are promising for future generations of GPUs, but they are not available on current GPUs, which motivates the need for software optimizations.
Moreover, hardware optimizations, if implemented, are potentially synergistic with the software optimizations we propose.

Many compiler/software optimizations have been proposed to improve dynamic parallelism performance or to provide an alternative to dynamic parallelism.
CUDA-NP~\cite{yang2014cudanp} is a compiler approach that enables annotation of parallel loops in the code with directives, instead of using dynamic parallelism.
The compiler transforms the kernel to launch excess threads for each original thread.
Using control flow, the excess threads are activated whenever a parallel loop is encountered.
A later work~\cite{khorasani2016eliminating} improves this technique by having multiple original threads share the excess threads for better load balance and less control divergence.
Another approach, Free Launch~\cite{chen2015free}, eliminates launches of child grids by applying transformations that reuse parent threads to execute child threads either sequentially or in parallel.
These approaches mitigate the overhead of dynamic parallelism by avoiding it entirely, but require threads to be on standby regardless of whether or not there is work available for them to do.

Li et al.~\cite{li2015exploiting,li2015nested}, Wu et al.~\cite{wu2016compiler}, and KLAP~\cite{el2016klap,el2018techniques} are aggregation techniques where multiple child grids are combined into a single aggregated grid to reduce the launch overhead.
Zhang et al.~\cite{zhang2018taming,zhang2018transforming} further enhance this approach by grouping together child grids with similar optimal configurations rather than placing all child grids in the same aggregated grid.
We leverage one of these works, KLAP~\cite{el2016klap}, as the aggregation component in our flow.

KLAP~\cite{el2016klap} also includes another dynamic parallelism optimization, promotion, which targets a specific pattern where a single-block kernel calls itself recursively.
Our optimizations are not applicable to this pattern.
Thresholding is not applicable because all child grids have the same size.
Coarsening is not applicable because a child grid has only one block.
Aggregation is not applicable because only one thread per parent grid performs a launch.

Various frameworks have been proposed to optimize the execution of applications with irregular parallelism on GPUs.
Wireframe~\cite{abdolrashidi2017wireframe}, Juggler~\cite{belviranli2018juggler}, ATA~\cite{helal2019adaptive}, and BlockMaestro~\cite{abdolrashidi2021blockmaestro} facilitate the execution of irregular parallel workloads where data-dependences between thread blocks need to be enforced.
VersaPipe~\cite{zheng2017versapipe} facilitates the extraction of pipeline parallelism from different GPU kernels.
NestGPU~\cite{floratos2021nestgpu} optimizes the execution of nested SQL queries on GPUs while avoiding the use of dynamic parallelism due it its inefficiency.
Our work focuses on optimizing GPU applications with nested parallelism expressed using dynamic parallelism.

%% file: sec/9-conclusion.tex
\section{Conclusion}\label{sec:conclusion}

We present an open-source compiler framework for optimizing the use of dynamic parallelism in applications with nested parallelism.
The framework includes three key optimizations: thresholding, coarsening, and aggregation.
Our evaluation shows that our compiler framework substantially improves performance of applications with nested parallelism that use dynamic parallelism, compared to when dynamic parallelism is not used or when it is used with aggregation only like in prior work.

\section*{Acknowledgments}

This work is supported by the University Research Board of the American University of Beirut (URB-AUB-103782-25509).

%% file: sec/appendix.tex
\appendix
\section{Artifact Appendix}

\subsection{Abstract}

Our artifact is a compiler for optimizing applications that use dynamic parallelism following the workflow illustrated in Figure~\ref{fig:framework}(a).
We have implemented the compiler in Clang~\cite{lattner2004llvm} and have made the compiler code publicly available.
Since building the compiler requires building Clang/LLVM which can be time and resource consuming, we provide pre-built binaries of the compiler in a Docker image, along with the required dependences and the benchmarks/datasets on which the compiler has been evaluated.
Reviewers can use the compiler binaries to transform the benchmark CUDA code with our optimizations, then compile and run the code on a CUDA-capable GPU to verify the timing/speedup results reported in Section~\ref{sec:evaluation}.
Scripts are provided to automate this process.

\subsection{Artifact check-list (meta-information)}
{\small
\begin{itemize}
    \item {\bf Program: }
        The benchmarks used are listed in Table~\ref{tab:benchmarks-and-datasets}.
        The benchmark code is obtained from prior work~\cite{el2016klap}\footnote{Prior work's code is available here: https://github.com/illinois-impact/klap}.
        We also include copies of the benchmark code in the artifact.
    \item {\bf Compilation:}
        The CUDA code for the benchmarks before and after transformation requires NVCC to be compiled, which has been included in the Docker image in the artifact.
    \item {\bf Transformations:}
        The software transformations are implemented in Clang~\cite{lattner2004llvm}.
        The code for these transformations is in the artifact, and a pre-compiled binary from this code has been included in the Docker image in the artifact.
    \item {\bf Binaries:}
        We include x86-x64 Linux binaries for our pre-compiled Clang-based compiler passes in the Docker image in the artifact.
    \item {\bf Data set:}
        The datasets are listed in Table~\ref{tab:benchmarks-and-datasets} and have been included in the artifact.
    \item {\bf Run-time environment:}
        The Docker image in the artifact builds x86-x64 Linux binaries and includes all the dependences.
    \item {\bf Hardware:}
        The evaluation requires a CUDA-cabable GPU that can execute dynamic parallelism code.
        Our evaluation used a V100 GPU with 32GB of memory.
        We recommend having at least 16GB of GPU memory to support the datasets.
    \item {\bf Metrics:}
        The metric reported is execution time/speedup.
    \item {\bf Experiments: }
        Two scripts are provided in the artifact:
            one script that tests the best configuration (threshold value, coarsening factor, aggregation granularity) for each combination of optimizations (thresholding, coarsening, aggregation) to verify Figures~\ref{fig:performance} and~\ref{fig:speedup-road-graphs};
            and one script that exhaustively tests all possible configurations for each combination of optimizations to verify Figure~\ref{fig:sweep-ta-multiblock}.
        In our evaluation, ten runs are used and an average is taken, but there is little variation across runs for most benchmarks/datasets.
        We do not include a script for reproducing the results in Figure~\ref{fig:breakdown} because the process of collecting these results (described in Section~\ref{sec:methodology}) is manual and difficult to automate.
    \item {\bf Output:}
        Running the binary of a single benchmark outputs the time it takes for that benchmark to run.
        Running the experiment scripts outputs a CSV file with the execution time of each benchmark/dataset for each combination of compiler optimizations/configurations used.
    \item {\bf How much disk space required (approximately)?:}
        $<$ 4GB
    \item {\bf How much time is needed to prepare the workflow (approximately)?:}
        $<$ 1hr
    \item {\bf How much time is needed to complete the experiments (approximately)?:}
        A single benchmark compilation and run should finish in $<$ 1min.
        Running the script for the best configurations for all benchmarks should take $<$ 1hr.
        Running the script for the exhaustive search can take up to 24hrs per benchmark.
    \item {\bf Publicly available?:}  Yes.
    \item {\bf Code licenses (if publicly available)?:} MIT License.
    \item {\bf Archived (provide DOI)?:} Yes, at the following link: https://doi.org/10.6084/m9.figshare.17048447.v1.
\end{itemize}
}

\subsection{Description}

\subsubsection{How to access}

The artifact can be downloaded at the following link:\\https://doi.org/10.6084/m9.figshare.17048447.v1.

The artifact consists of the code for the compiler passes, the benchmark code, the datasets, and a Docker image which includes the binaries of the compiler passes and the dependences required to execute the compiler passes, compile the transformed benchmark code, and execute the benchmarks.

\subsubsection{Hardware dependencies}

Running the binaries depends on having a CUDA capable devices.
Our evaluation used a V100 GPU with 32GB of memory.
We recommend having at least 16GB of GPU memory to support the datasets.

\subsubsection{Software dependencies}

We ran our tests on an environment with CUDA-9.1 installed.
Software dependencies have been packaged in the Docker image in the artifact.

\subsubsection{Data sets}

The datasets are listed in Table~\ref{tab:benchmarks-and-datasets} and have been included in the artifact.


\subsection{Installation}

The artifact contains a README file with installation instructions and a convenience script for handling the installation.


\subsection{Experiment workflow}

To run the experiments, run the Docker image: `./run.sh`.

In order to compile all the binaries with default parameters, from inside the Docker container, change directory to the relevant benchmark inside the `test/` directory and run `make all`.

In order to run the benchmark with the best configurations, run `bestcombination.sh`.
The parameters used here are the best combination of parameters we found for each benchmark and data set when we exhaustively searched the space.

In order to perform the exhaustive search, run `sweep.sh` which is available inside every benchmark directory.

\subsection{Evaluation and expected result}

Running `bestcombination.sh` for each benchmark should provide the execution times used to report the speedups in Figures~\ref{fig:performance} and~\ref{fig:speedup-road-graphs}.
Running `sweep.sh` for each benchmark should provide the execution times used to report the speedups in Figure~\ref{fig:sweep-ta-multiblock}.
In our evaluation, ten runs are used and an average is taken, but there is little variation across runs for most benchmarks/datasets.


\subsection{Experiment customization}

The experiments can be customized by running each benchmark with the desired parameters.
The parameters can be updated in the provided Makefile for each benchmark.




